\documentclass[prl,twocolumn,showpacs,floatfix,superscriptaddress,%
preprintnumbers,nofootinbib]{revtex4}

\usepackage{graphicx,color,natbib}
\usepackage{amsmath,bm}

\begin{document}

 \title{Suppression of Self-Induced Flavor Conversion in the Supernova Accretion Phase}

\author{Srdjan~Sarikas}
\affiliation{Dipartimento di Scienze Fisiche, Universit\`a di Napoli
Federico~II,
80126~Napoli, Italy}

\affiliation{Max-Planck-Institut f\"ur Physik
(Werner-Heisenberg-Institut), F\"ohringer Ring~6, 80805
          M\"unchen, Germany}

\author{Georg~G.~Raffelt}
\affiliation{Max-Planck-Institut f\"ur Physik
(Werner-Heisenberg-Institut), F\"ohringer Ring~6, 80805
          M\"unchen, Germany}

\author{Lorenz~H\"udepohl} \affiliation{Max-Planck-Institut f\"ur
  Astrophysik, Karl-Schwarzschild-Str.~1, 85748 Garching, Germany}

\author{Hans-Thomas~Janka} \affiliation{Max-Planck-Institut f\"ur
  Astrophysik, Karl-Schwarzschild-Str.~1, 85748 Garching, Germany}

\date{15 September 2011}

\begin{abstract}
Self-induced flavor conversions of supernova (SN) neutrinos can
strongly modify the flavor dependent fluxes. We perform a linearized
flavor stability analysis with accretion-phase matter profiles of a
$15\,M_\odot$ spherically symmetric model and corresponding neutrino
fluxes. We use realistic energy and angle distributions, the latter
deviating strongly from quasi-isotropic emission, thus accounting
for both multi-angle and multi-energy effects. For our matter and
neutrino density profile we always find stable conditions: flavor
conversions are limited to the usual Mikheyev-Smirnov-Wolfenstein effect. In this case one
may distinguish the neutrino mass hierarchy in a SN neutrino signal
if the mixing angle $\theta_{13}$ is as large as suggested by recent
experiments.
\end{abstract}

\preprint{MPP-2011-110}

\maketitle

{\em Introduction.}---The huge neutrino fluxes emitted by
core-collapse supernovae (SNe) are key to the explosion dynamics and
nucleosynthesis~\cite{Janka:2006fh} and detecting a high-statistics
``neutrino light curve'' from the next nearby SN is a major goal for
neutrino astronomy~\cite{Autiero:2007zj}. Besides probing the
core-collapse phenomenon in unprecedented detail, one may detect
signatures of flavor oscillations and extract information on neutrino
mixing parameters~\cite{Dighe:1999bi, Chakraborty:2011ir}.

The refractive effect caused by matter~\cite{Wolfenstein:1977ue}
suppresses flavor oscillations until neutrinos pass through the Mikheyev-Smirnov-Wolfenstein (MSW)
region in the collapsing star's envelope~\cite{Mikheev:1986gs,Kuo:1989qe}. However, neutrino-neutrino interactions, through a
flavor off-diagonal refractive index~\cite{Pantaleone:1992eq,Sigl:1992fn}, can trigger self-induced flavor
conversions~\cite{Samuel:1993uw,Sawyer:2005jk,Sawyer:2008zs}. This
collective effect usually occurs between the neutrino sphere and the
MSW region and can strongly modify neutrino spectra
\cite{Duan:2006an, Dasgupta:2009mg, Duan:2010bg}, although this would
never seem to help explode the star \cite{Dasgupta:2011jf}. Actually,
in low-mass SNe (not studied here) the density falls off so fast that
MSW can occur first, leading to novel effects on the prompt $\nu_e$
burst~\cite{Duan:2007sh}.

Collective oscillations at first seemed unaffected by matter because
its influence does not depend on neutrino energies
\cite{Duan:2006an}. However, depending on emission angle, neutrinos
accrue different matter-induced flavor-dependent phases until they
reach a given radius. This ``multi-angle matter effect'' can
suppress self-induced flavor conversion~\cite{EstebanPretel:2008ni}.
Based on schematic flux spectra, this was numerically confirmed for
accretion-phase SN models where the density near the core is
large~\cite{Chakraborty:2011gd}. This epoch, before the delayed
explosion finally takes off, is when the neutrino luminosity and the
difference between the $\bar\nu_e$ and $\bar\nu_{\mu,\tau}$ fluxes
are largest. If self-induced flavor conversion did not occur and the
mixing angle $\theta_{13}$ was not very small \cite{Fogli:2011qn},
the accretion phase would provide a plausible opportunity to
determine the mass hierarchy~\cite{Dighe:1999bi,Chakraborty:2011gd}.

Numerical multi-angle simulations of collective oscillations are very
demanding~\cite{Duan:2008eb}, but it is much easier to study if such
oscillations are suppressed for given density profile and neutrino
distributions. Self-induced conversion requires that part of the
spectrum is prepared in one flavor, the rest in another. The
collective mode consists of pendulum-like flavor exchange between
these parts without changing the overall flavor content
\cite{Samuel:1993uw,Hannestad:2006nj}. The inevitable starting point
is a flavor instability of the neutrino distribution caused by
neutrino-neutrino refraction. An exponentially growing mode can be
detected with a linearized analysis of the evolution
equations~\cite{Sawyer:2008zs,Banerjee:2011fj}. We here apply this
method to a numerical accretion-phase SN model, for the first time
using both realistic neutrino energy spectra and angular
distributions.

\begin{figure}
\includegraphics[width=0.75\columnwidth]{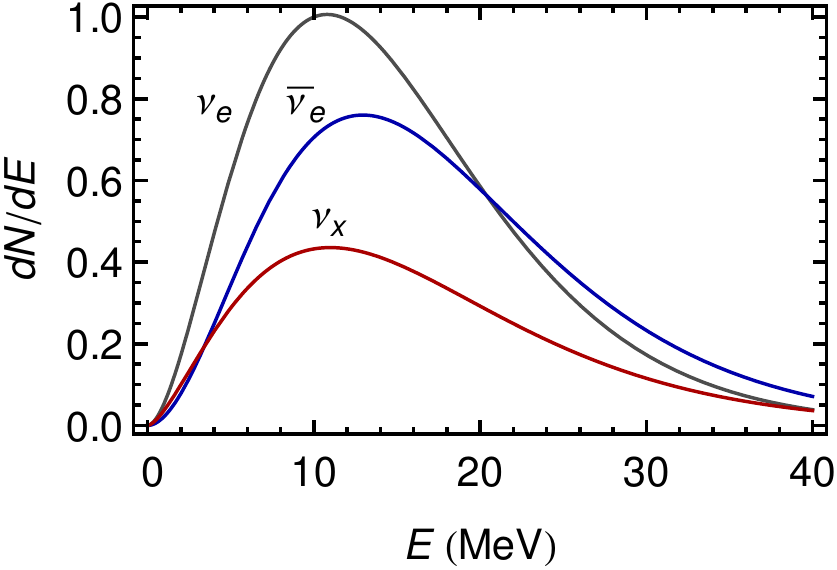}
\vskip 10pt
\includegraphics[width=0.75\columnwidth]{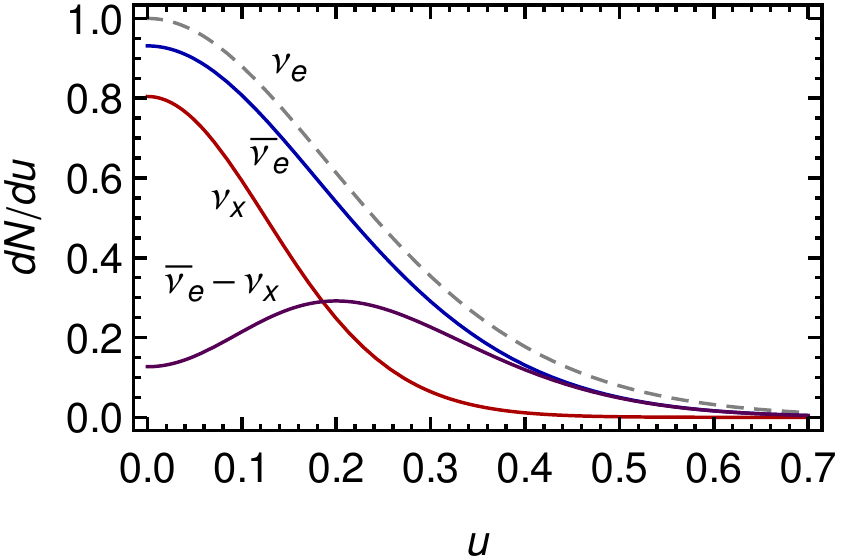}
\caption{Flux spectra for our 280~ms SN model.
The angle variable $0\leq u\leq1$ is based on $R=44.7$~km.
\label{fig:fluxes}}
\end{figure}

{\em Numerical SN model.}---Our spherically symmetric simulation of
an accretion-phase SN model was performed with the {\sc
Prometheus-Vertex} code as in Ref.~\cite{Huedepohl:2009wh}, now with
a $15\,M_\odot$ progenitor~\cite{Woosley:1995ip}. The transport
module computes the energy and angle distributions of $\nu$ and
$\bar\nu$ for all flavors by a tangent-ray discretization of the
Boltzmann transport equation~\cite{Rampp:2002bq}. We used 21 nearly
geometrically spaced energy bins up to 380 MeV and 672 tangent rays.
We do not artificially trigger an explosion, but otherwise our model
is comparable to Ref.~\cite{Chakraborty:2011gd}. We use several
snapshots and illustrate our findings with one at 280~ms post bounce.
The flux spectra (Fig.~\ref{fig:fluxes}) show a $\nu_e$ excess from
deleptonization and a $\bar\nu_e$ flux almost twice that of $\nu_x$
(representing any of $\nu_{\mu,\tau}$ or $\bar\nu_{\mu,\tau}$) and
average $\nu_e$, $\bar\nu_e$ and $\nu_x$ energies of 15.3, 18.1, and
16.9~MeV.

We study neutrino propagation in the free-streaming limit, so we can
describe the angular distribution by the angle $\vartheta_R$
relative to the radial direction at a chosen inner-boundary radius
$R$. Actually it is more convenient to use
$u=\sin^2\vartheta_R=(1-\cos^2\vartheta_r)\,r^2/R^2$, which is
uniformly distributed on $0\leq u\leq1$ if emission is isotropic at
a ``neutrino sphere'' with radius $R$ \cite{EstebanPretel:2008ni,Banerjee:2011fj}. We choose $R=44.7$~km and show the corresponding
$u$ distribution in Fig.~\ref{fig:fluxes}. Isotropic emission from a
neutrino sphere is not a good description because neutrinos emerge
from a thick layer. The $\bar\nu_e$ and $\nu_x$ intensities are
similar in the radial direction: the excess $\bar\nu_e$ flux largely
arises from its broader angular distribution (larger emission
region). Flavor oscillations depend on the difference of the $e$ and
$x$ distributions, which is small in the radial direction
(Fig.~\ref{fig:fluxes}). The angular distributions do not cross,
although in principle there could have been a forward $\nu_x$
excess.

\begin{figure}
\includegraphics[width=0.72\columnwidth]{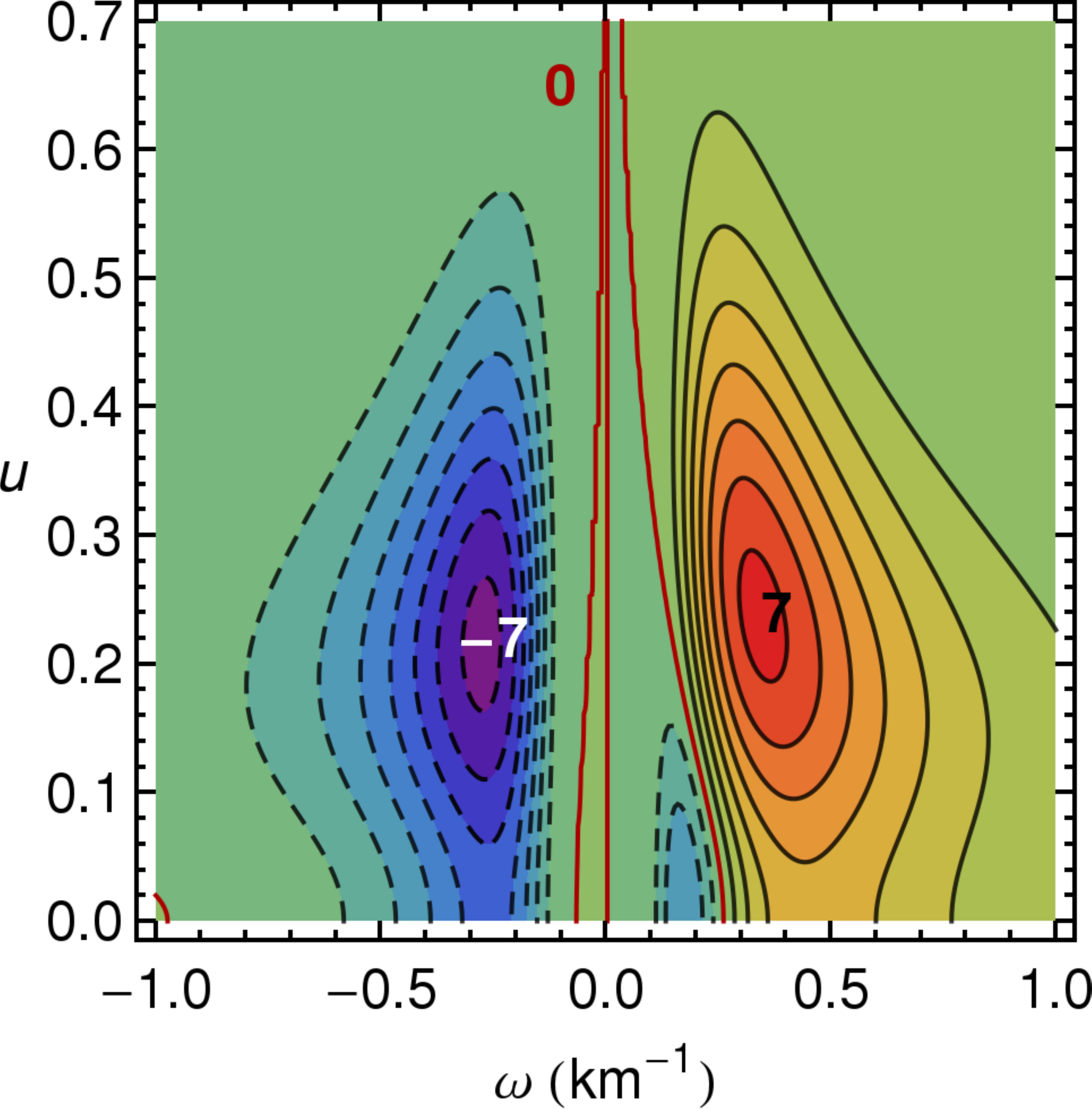}
\caption{Distribution $g(\omega,u)$ describing the neutrino fluxes.\label{fig:g_distribution}}
\end{figure}

In the context of neutrino oscillations, $\omega=\Delta m^2/2E$ is a
preferred energy variable, where $\Delta m^2=(50~{\rm meV})^2$ is
the ``atmospheric'' neutrino mass-squared difference relevant for
1--3 oscillations studied here. Moreover, treating anti-neutrinos
formally as negative-energy neutrinos with negative occupation
numbers vastly simplifies the formalism. Flavor oscillations can
exchange $\nu_e$ with $\nu_x$, leaving the overall neutrino flux
unchanged, so only $F_{\nu_e}{-}F_{\nu_x}$ matters. Our sign
convention means that for anti-neutrinos we then use
$F_{\bar\nu_x}-F_{\bar\nu_e}$, corresponding to the flavor isospin
convention~\cite{Duan:2006an}. The neutrino flux difference
distribution $g(\omega,u)$ thus defined is shown in
Fig.~\ref{fig:g_distribution}. It is negative for anti-neutrinos
($\omega<0$) because $F_{\bar\nu_e}>F_{\bar\nu_x}$. For
$\omega\sim0.2~{\rm km}^{-1}$ there is a spectral crossing as a
function of $u$, i.e.\ for large $E$ the $\nu_x$ flux does exceed
the $\nu_e$ flux in the forward direction.

Self-induced oscillations exchange the positive and negative parts of
$g(\omega,u)$, leaving fixed the overall flavor content
$\varepsilon=(F_{\nu_e}{-}F_{\nu_x})/(F_{\bar\nu_e}{-}F_{\bar\nu_x})-1=\int
d\omega d u g(\omega,u)$. Our $g(\omega,u)$ is mostly negative for
$\bar\nu$ and mostly positive for $\nu$, so collective oscillations
largely correspond to pair conversions
$\nu_e\bar\nu_e\leftrightarrow\nu_x\bar\nu_x$. Accretion-phase
distributions are ``single crossed'' in this sense, i.e.\
$g(\omega,u)$ changes sign essentially only on the line $\omega=0$,
because of the large excess of the  $\nu_e$ and $\bar\nu_e$ fluxes.
Significant multiple crossings are typical for the cooling
phase~\cite{Dasgupta:2009mg}.

{\it Equations of motion (EoM).}---We describe three-flavor neutrino
propagation by energy- and angle-dependent $3{\times}3$ matrices
${\bf\Phi}_{E,u}(r)$. Boldface characters denote matrices in flavor
space. The diagonal ${\bf\Phi}_{E,u}$ elements are the ordinary
number fluxes $F_{E,u}^\alpha$ (flavor $\alpha$) integrated over a
sphere of radius $r$, with negative $E$ and negative number fluxes for
anti-neutrinos. The off-diagonal elements, which are initially zero,
represent phase information caused by flavor oscillations. The
flavor evolution is then provided by the ``Schr\"odinger equation''
${\rm i}\partial_r{\bf\Phi}_{E,u}=[{\bf H}_{E,u},{\bf\Phi}_{E,u}]$
with the Hamiltonian \cite{EstebanPretel:2008ni}
\begin{eqnarray}\label{eq:hamiltonian}
{\bf H}_{E,u}&=&\frac{1}{v_{u}}\,\left(\frac{{\bf M}^2}{2E}+
\sqrt{2}\,G_{\rm F}{\bf N}_\ell\right)\\
&+&\frac{\sqrt{2}\,G_{\rm F}}{4\pi r^2}\int_{-\infty}^{+\infty}dE'
\int_0^1 du'
\frac{1-v_{u}v_{u'}}{v_{u}v_{u'}}\,{\bf\Phi}_{E',u'}\,.
\nonumber
\end{eqnarray}
The matrix ${\bf M}^2$ of neutrino mass-squares causes vacuum flavor
oscillations and that of net charged lepton densities ${\bf
N_\ell}={\rm diag}(n_e{-}n_{\bar
e},n_\mu{-}n_{\bar\mu},n_\tau{-}n_{\bar\tau})$ adds the Wolfenstein
matter effect. The third term provides neutrino-neutrino refraction
and is analogous to matter except for Pantaleone's off-diagonal
elements and except that in the SN context neutrinos are not
isotropic. A neutrino radial velocity at radius $r$ is
$v_u=(1-u\,R^2/r^2)^{1/2}$. The factor $1-v_u v_{u'}$ arises from
the current-current nature of the weak interaction and causes
multi-angle effects. Moreover, $v_u$ appears in the denominator
because we follow the flavor evolution projected on the radial
direction, causing the multi-angle matter
effect~\cite{EstebanPretel:2008ni}.

Up to the MSW region, the matter effect is so large that
${\bf\Phi}_{E,u}$ is very nearly diagonal in the weak-interaction
basis, the usual approximation made in SN neutrino transport.
Neutrinos remain stuck in flavor eigenstates unless the off-diagonal
${\bf\Phi}_{E,u}$ elements start growing by the self-induced
instability. To find the latter we linearize the EoM in the small
off-diagonal amplitudes.

{\it Stability condition.}---We study the instability driven by the
atmospheric $\Delta m^2$ and the mixing angle $\theta_{13}\ll1$, we
work in the two-flavor limit, and switch to the $\omega=\Delta
m^2/2E$ variable. We write the flux matrices in the form
\begin{equation}
{\bf\Phi}_{\omega,u}=\frac{{\rm Tr}\,{\bf\Phi}_{\omega,u}}{2}
+\frac{F_{\omega,u}^e-F_{\omega,u}^x}{2}\,
\begin{pmatrix}
s_{\omega,u}&S_{\omega,u}\\
S_{\omega,u}^*&-s_{\omega,u}
\end{pmatrix}\,,
\end{equation}
where $F_{\omega,u}^{e,x}$ are the flavor fluxes at the inner
boundary radius $R$. The flux summed over all flavors, ${\rm
Tr}\,{\bf\Phi}_{\omega,u}$, is conserved in our free-streaming
limit. The $\nu_e$ survival probability is
$\frac{1}{2}[1+s_{\omega,u}(r)]$ in terms of the ``swap factor''
$-1\leq s_{\omega,u}(r)\leq1$. The off-diagonal element
$S_{\omega,u}$ is complex and $s^2_{\omega,u}+|S_{\omega,u}|^2=1$.

The small-amplitude limit means $|S_{\omega,u}|\ll1$ and to linear
order $s_{\omega,u}=1$. Assuming in addition a large distance from
the source so that $1-v_u\ll 1$, the evolution equation linearized in
$S_{\omega,u}$ and in $u$ is \cite{Banerjee:2011fj}
\begin{eqnarray}\label{eq:smallEoM}
{\rm i}\partial_r S_{\omega,u}&=&
\left[\omega+u(\lambda+\varepsilon\mu)\right]S_{\omega,u}
\nonumber\\
& &-\mu \int du'\,d\omega'\,(u+u')\,g_{\omega'u'}\,S_{\omega',u'}\,.
\label{stability-eom}
\end{eqnarray}
Weak-interaction effects are encoded in the $r$-dependent parameters
with dimension energy
\begin{eqnarray}
\lambda&=&\sqrt{2}\,G_{\rm F}\,[n_e(r)-n_{\bar e}(r)]\,\frac{R^2}{2r^2}\,,
\nonumber\\
\mu&=&\frac{\sqrt{2}\,G_{\rm F}\,[F_{\bar\nu_e}(R)-F_{\bar\nu_x}(R)]}{4\pi r^2}
\,\frac{R^2}{2r^2}\,.
\end{eqnarray}
The factor $R^2/2r^2$ signifies that only the multi-angle impact of 
the $\nu$-$\nu$ and matter effects are relevant for the stability analysis, not the densities themselves. Both $\lambda$ and
$\mu$ depend on $R$, but so does the occupied $u$-range: physical
results do not depend on the choice of~$R$. We choose $R=44.7$~km such that the
occupied $u$-range is 0--1. We normalize the $\nu$-$\nu$ interaction
strength $\mu$ to the $\bar\nu_e$-$\bar\nu_x$ flux difference at
$R$, i.e.\ $\int_{-\infty}^0 d\omega\int_0^1 du\,g_{\omega,u}=-1$,
but the only physically relevant quantity is $\mu(r)\,g_{\omega,u}$.
Our SN model provides $\mu(R)=1.73 \times 10^{4}~{\rm km}^{-1}$ and
an ``asymmetry''  $\varepsilon=\int d\omega\,
du\,g_{\omega,u}=0.35$.

Writing solutions of the linear differential equation,
Eq.~(\ref{eq:smallEoM}), in the form
$S_{\omega,u}=Q_{\omega,u}\,e^{-{\rm i}\Omega r}$ with complex
frequency $\Omega=\gamma+{\rm i}\kappa$ and eigenvector
$Q_{\omega,u}$ leads to the eigenvalue equation
\cite{Banerjee:2011fj},
\begin{equation}\label{fourier-eom}
(\omega + u \bar\lambda - \Omega)\, Q_{\omega,u}=
\mu \int du'\,d\omega'\,(u+u')\,g_{\omega'u'}\,Q_{\omega',u'}\,,
\end{equation}
where $\bar\lambda \equiv \lambda + \varepsilon \mu$. The solution
has to be of the form
$Q_{\omega,u}=(A+Bu)/(\omega+u\bar\lambda-\Omega)$. Solutions exist
if $\mu^{-1}=I_1\pm\sqrt{I_0I_2}$, where $I_n=\int
d\omega\,du\,g_{\omega,u}\,u^n/(\omega+u\bar\lambda-\Omega)$. The
system is stable if all $\Omega$ are purely real. A possible
imaginary part, $\kappa$, is the exponential growth rate.

\begin{figure}
\includegraphics[width=0.78\columnwidth]{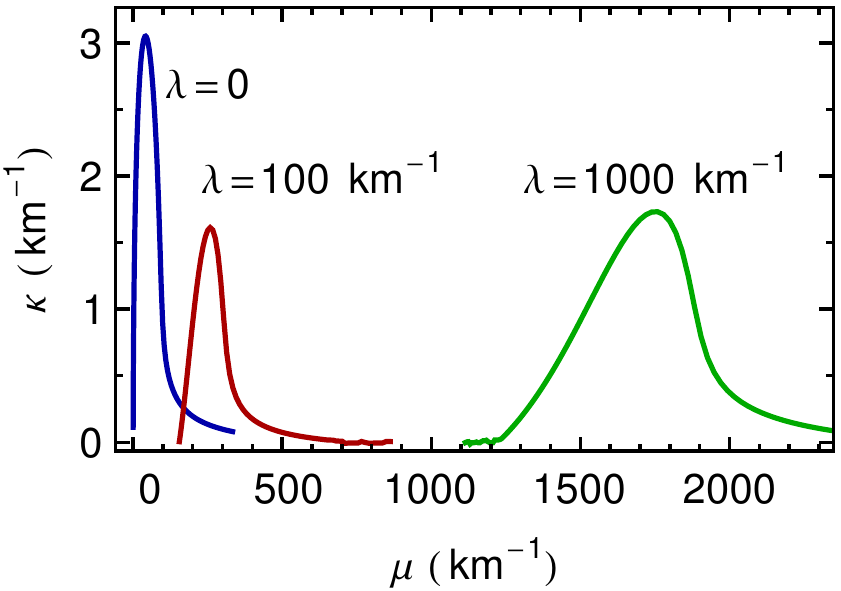}
\caption{Growth rate $\kappa$ for our SN model as a function of $\mu$
for various $\lambda$ values as indicated.\label{fig:kappa}}
\end{figure}

{\it Application to our SN model.}---Ignoring the effect of matter
($\lambda=0$), we show $\kappa(\mu)$ for our 280~ms SN model in
Fig.~\ref{fig:kappa}. The system is essentially stable above $\mu$
of a few 100~km$^{-1}$. It is noteworthy that $\kappa$ is of the
same order as a typical $\omega$ of the $g_{\omega,u}$ distribution,
in our case a few km$^{-1}$. We also show $\kappa(\mu)$ for
$\lambda=10^2$ and $10^3$~km$^{-1}$ and observe a shift to larger
$\mu$-values~\cite{Banerjee:2011fj}.

\begin{figure}[b]
\includegraphics[width=0.78\columnwidth]{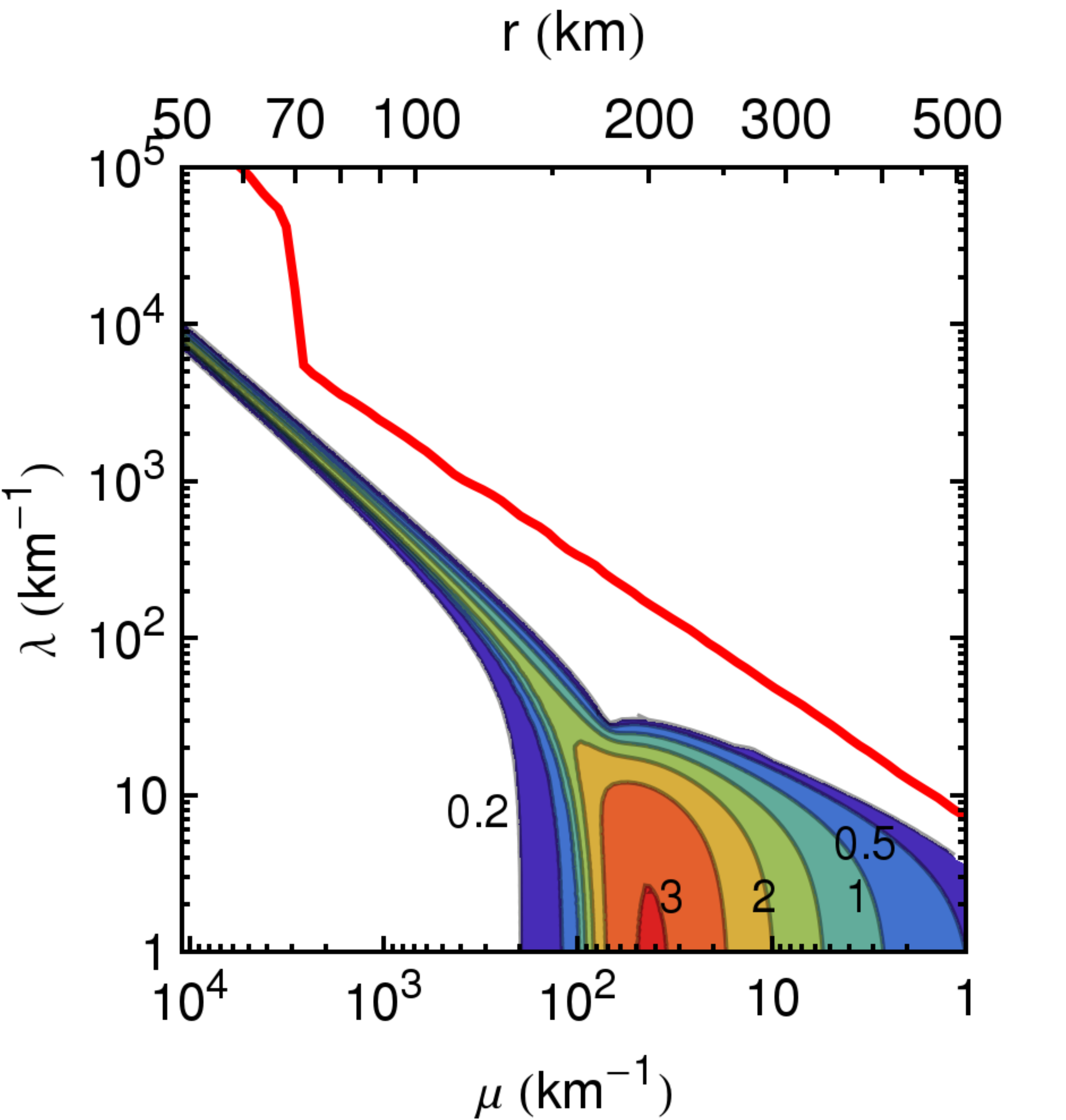}
\caption{Contours for the growth rate $\kappa$ in
km$^{-1}$. Also shown is the profile for our
SN model. The vertical axis essentially denotes the density,
the horizontal axis the radius ($\mu\propto r^{-4}$).
\label{fig:trajectory}}
\vskip-12pt
\end{figure}

In Fig.~\ref{fig:trajectory} we show contours of $\kappa$ in the
$(\mu,\lambda)$ plane. For large $\mu$ and $\lambda$ values, the
system is unstable for $\lambda\sim\mu$ \cite{Banerjee:2011fj}. In
other words, if the local neutrino number density is much smaller or
much larger than the local electron density, the system is stable.

We also show the locus of $[\mu(r),\lambda(r)]$ along the radial
direction. Since $\mu(r)\propto r^{-4}$, the red solid line in
Fig.~\ref{fig:trajectory} is essentially the SN density profile. The
step-like feature is the shock wave where the matter density drops by
an order of magnitude. Without matter ($\lambda=0$), neutrinos would
enter the instability strip at $\mu\sim100$, corresponding to $r\sim
150$~km. We find similar results for other snapshots at times 150 and
400~ms postbounce, i.e., neutrinos do not encounter a self-induced
instability.

{\it Comparison with earlier work.}---A similar accretion-phase model
of the Basel group was used to study flavor stability by numerically
solving the EoMs~\cite{Chakraborty:2011gd}. A mono-energetic neutrino
distribution and isotropic emission at a neutrino sphere were
assumed. For some snapshots, an instability occurred at a large
radius. Applying our method to the same matter profile and schematic
neutrino distribution we find perfect agreement with
Ref.~\cite{Chakraborty:2011gd} and even reproduce their onset radius
for those cases where partial flavor conversion
occurs~\cite{Sarikas:2011jc}. It would be interesting to repeat our
study with realistic Basel distributions to see if partial flavor
conversion is an artifact of their schematic energy and angle
distributions.

{\it Conclusions.}---We have performed a linearized flavor stability
analysis of accretion-phase SN models and neutrino fluxes with
realistic energy and angle distributions. For these models,
self-induced flavor conversions do not occur. One should apply this
method to a broader class of models to see if this conclusion is
generic. It also remains to extend a linearized analysis to cases
without cylindrical symmetry of the angular distribution in view of
Sawyer's concerns about a significant multi-angle
instability~\cite{Sawyer:2008zs}. In realistic 3D models, the
neutrino distribution is not cylindrically symmetric and even if this
were the case, in principle even a small fluctuation could trigger a
novel instability if it were to exist.

Recent evidence suggests that the neutrino mixing angle $\theta_{13}$
is not very small~\cite{Fogli:2011qn}, implying that the MSW
conversion in the SN is adiabatic. One can then distinguish the
neutrino mass hierarchy by Earth matter effects~\cite{Dighe:1999bi}
or the early rise time~\cite{Chakraborty:2011ir}, but only if
collective oscillations do not swap flavors before the MSW region.
(In the presence of the collective flavor swap, the hierarchy could
be distinguished for an extremely small $\theta_{13}$ where the MSW
conversion is not adiabatic~\cite{Dasgupta:2008my}.) The suppression
of self-induced flavor conversion during the accretion phase, if
generic, is good news for the possibility of determining the mass
hierarchy with SN neutrinos if a ``large'' value for $\theta_{13}$ is
experimentally confirmed.


\begin{acknowledgments}
We thank T.~Hahn for helping to implement a numerical
library~\cite{Hahn:2004fe}. We acknowledge partial support by DFG
Grants No.\ TR~7, TR~27, EXC~153 and computer time at the HLRS in
Stuttgart and NIC in J\"ulich.
\end{acknowledgments}


\end{document}